\documentclass[twocolumn,secnumarabic,amssymb,
amsmath,aps,prl,showpacs,showkeys]{revtex4}
\usepackage{graphics}
\usepackage{graphicx}
\begin{document}

\title{Fourier's law of heat conduction in a three dimensional 
harmonic crystal: A retrospection} 

\author{ Shiladitya Acharya \altaffiliation{Permanent Address:}
and  Krishnendu Mukherjee}
\affiliation{Department of Physics, Bengal Engineering and Science University,
Shibpur, Howrah- 711 103, West Bengal, India}
\email{kmukherjee@physics.becs.ac.in}

\begin{abstract}
We present an exact solution of the Langevin's equation in the 
steady state limit in a three dimensional, harmonic crystal of
slab geometry whose boundary surfaces along its length are 
connected to two stochastic, white noise heat baths at different 
temperatures. We show that the heat transport obeys the Fourier's
law in the continuum limit. 
\end{abstract}

\maketitle

\vskip 2in
When a steady temperature gradient is established 
between the two ends of a piece of solid bar, heat
current will flow from high to low temperature end. 
According to Fourier's law of heat conduction the current 
density is proportional to the temperature gradient and 
mathematically it reads as
\begin{equation}
 {\bf J}({\bf x})=-\kappa\nabla T({\bf x}),  
\label{Fourier}
\end{equation}  
where the constant of proportionality $\kappa$ is known as the 
thermal conductivity of the solid. Conduction of heat in solid 
by its very nature is a non-equilibrium process. This is
an area of Physics, where the idea of non-equilibrium statistical
mechanics can be applied to in order to find the underlying physical 
conditions for the validity of this law in case of solid.  
Various numerical and analytical studies confirm that 
the heat transport in one dimensional system exhibits 
anomalous\cite{Lepri1} behaviour. It means that the thermal conductivity
for such a system is not found to be an intrinsic property of the
material. It shows a power law dependence $\kappa \sim N^{\alpha}$,
where $N$ be the linear size of the system.
There are studies on different models which predict divergent 
($0<\alpha<1$) thermal conductivity
\cite{Lepri1,Dhar1,Grassberger1,Narayan,Mai}. 
There are also some oscillator models that give non-divergent ($\alpha <0$)
thermal conductivity\cite{Dhar2} in one dimension. 
The anomalous behaviour of thermal conductivity is also 
observed in two dimensional system.
Numerical study indicates a logarithmic divergence\cite{Lippi} 
of thermal conductivity $\kappa\sim \ln{N}$. 
A power law behaviour\cite{Grassberger2} is also observed in 
such a system.

There are strong numerical evidences\cite{Dhar3} that indicate the validity
of Fourier's law of heat conduction in one and two dimensional
systems with pining and anharmonicity. An extensive investigation
on heat transport in a three dimensional disordered harmonic
crystal has been carried out recently\cite{Chaudhuri}. 
The numerical simulation indicates the normal transport
of heat when this system is subjected to an external pining
potential. Though it is not been verified numerically, 
but a finite conductivity is predicted for this 
disordered system from analytical arguments. 
A more recent simulation 
study \cite{Dhar4} establishes for the first time the validity of
this law in three dimensional anharmonic crystal.
It thus also establishes the fact that the process of heat
conduction in three dimensional geometry is diffusive in nature.
Apart from bringing in a temperature dependent contribution to
the thermal conductivity, which is indeed the case for real
systems, it is confirmed that anharmonicity 
provides a condition which is sufficient for normal heat 
transport in a solid.
In this letter we give an exact analytical
derivation of Fourier's law of heat conduction
in three dimensional harmonic crystal. 
We find that in the continuum limit
the thermal conductivity is finite and does not depend on the system size. 

We consider a cubic crystal in three dimension.
The form of the Hamiltonian  
\begin{eqnarray}
H = \sum_{\bf n} \frac{\dot x_{\bf n}^2}{2}+\sum_{\bf n,\hat{e}} \frac{1}{2} 
 (x_{\bf n} - x_{\bf n+\hat{e}})^2. 
\end{eqnarray} 
The displacement field $x_{\bf n}$ is 
defined on each lattice site
${\bf n}=(n_1,n_2,n_3)$ where $n_1=1,\cdots,N$, $n_2=1,\cdots,W_2$,
and $n_3=1,\cdots,W_3$.
Here ${\bf \hat{e}}$ denotes the unit vector in the three directions. 
We choose the value of mass attached to each lattice point and the 
harmonic spring constant as one.  
We have Langevin's type heat baths that are coupled to the 
surfaces at $n_1=1$ and $n_1=N$ and are maintained at temperatures
$T_L$ and $T_R$ ($T_L\,>T_R$) respectively. 
Hence the equation of motion of a particle at the site ${\bf n}$ reads 
\begin{eqnarray}
{\ddot{x}}_{\bf n}=&-&\sum_{{\bf \hat{e}}} (x_{\bf n}-x_{\bf n+\hat{e}})-
\gamma (\delta_{n_1,1}+ \nonumber
\delta_{n_1,N}){\dot{x}}_{\bf n}\\
&+&(\delta_{n_1,1} \eta_{\bf n}^L
+\delta_{n_1,N} \eta_{\bf n}^R).
\label{eom}
\end{eqnarray}
We have chosen the noises to be white and they are uncorrelated at
different sites. Noise strength is specified by
\begin{equation}
\langle \eta_{\bf n}^{L,R}(t)
\eta_{\bf n^\prime}^{L,R}(t)\rangle
=2\gamma T_{L,R}\delta(t-t^\prime)\delta_{{\bf{n}},{\bf{n}}^\prime},
\label{noise1}
\end{equation}
where we have chosen the Boltzmann constant $k_B=1$.
We use the periodic boundary conditions for the displacement 
field and the noises in $n_2$ and $n_3$ directions:
\begin{eqnarray} 
x_{{\bf n}+(0,W_2,0)}(t) &=& x_{\bf n}(t) = x_{{\bf n}+(0,0,W_3)}(t)\nonumber\\
\eta_{{\bf n}+(0,W_2,0)}^{L,R}(t) &=& \eta_{\bf n}^{L,R}(t)
=\eta_{{\bf n}+(0,0,W_3)}^{L,R}(t)
\label{pbc}  
\end{eqnarray}
These periodic boundary conditions lead to the following 
expansion of $x_{\bf n}(t)$ and $\eta_{\bf n}^{L,R}(t)$:  
\begin{eqnarray}
x_{\bf n}(t)&=&\frac{1}{\sqrt{W_2 W_3}}
\sum_{p_2} \sum_{p_3} y_{n_1}(p_2,p_3,t)
{\rm e}^{i(p_2 n_2+p_3 n_3)a},\nonumber\\
& & \label{ftxn}\\
\eta_{\bf n}^{L,R}(t)&=&\frac{1}{\sqrt{W_2 W_3}}
\sum_{p_2} \sum_{p_3} f_{n_1}(p_2,p_3,t)
{\rm e}^{i(p_2 n_2+p_3 n_3)a},\nonumber\\
& &\label{ftetan}
\end{eqnarray} 
where $a$ be the lattice constant of the crystal.
Upon substitution of Eqn.(\ref{ftxn}) and (\ref{ftetan}) into
Eqn.(\ref{eom}) we obtain 
\begin{equation}
\ddot{y}_j=-V_{jk}y_k-\gamma W_{jk}\dot{y}_k+f_j
\label{eom1}
\end{equation}
where  
\begin{eqnarray}
W_{jk} &=& \delta_{j,1}\delta_{k,1}
+\delta_{j,N}\delta_{k,N},\label{Wjk}\\
f_j(p_2,p_3,t) &=& \delta_{j,1}f_L(p_2,p_3,t)
+\delta_{j,N}f_R(p_2,p_3,t)\label{fj},
\end{eqnarray}
the $N\times N$ matrix 
\begin{equation}
V=\left(\begin{array}{ccccc}
2\omega_0^2 & -1 & 0 & 0 & \ldots  \\ 
-1 & 2\omega_0^2 & -1 & 0 & \ldots  \\
0 & -1 & 2\omega_0^2 & -1 & \ddots \\
\vdots & \ddots & \ddots & \ddots & \ddots  \\ 
0 & \ldots & 0 & -1 & 2\omega_0^2
\end{array}\right)
\end{equation}
and 
\begin{equation} 
\omega_0^2(p_2,p_3) = 1+2\sin^2(\frac{p_2 a}{2})
+2\sin^2(\frac{p_3 a}{2}).
\label{w0}
\end{equation} 
Here $j, k=1,\cdots, N$.
We have also assumed here
that $y_0(p_2,p_3,t)=0=y_{N+1}(p_2,p_3,t)$.
To solve Eqn.({\ref{eom1}}) we diagonalize the matrix $V$.
The solution of the $N$ order equation 
$\left|V-\alpha^2 I\right|=0$
gives the eigenvalues of $V$ as 
\begin{equation}
\alpha_k^2(p_1,p_2)=2\omega_0^2(p_1,p_2)+
2\cos\left(\frac{k\pi}{N+1}\right).
\label{eval} 
\end{equation}
The $j$-th component of the normalized eigenvector corresponding 
to the eigenvalue $\alpha_k^2$ is given by 
\begin{equation}
a^{(k)}_j=\sqrt{\frac{2}{N+1}} (-1)^{j+1} \sin\left(\frac{jk\pi}{N+1}\right).
\label{evec}
\end{equation} 
The diagonalizing matrix $A$ thus reads as
$A_{jk}=a^{(k)}_j$ such that $A^TA=I$ and 
$A^TVA=\alpha^2$, where $(\alpha^2)_{jk}=\alpha_j^2\,\delta_{jk}$.
We introduce a new set of coordinates $\xi_j$ as
\begin{equation}
y_j(p_2,p_3,t)=A_{jk}\xi_k(p_2,p_3,t).
\label{yxi}
\end{equation}
The equation of motion of $\xi_j$ in matrix form can be written as 
\begin{equation}
\ddot{\xi}=-\alpha^2\xi-\gamma Z \dot{\xi} +\tilde{f},
\label{eomxi} 
\end{equation}
where the symmetric matrix $Z=A^TWA$, and $\tilde{f}=A^Tf$. 
In the steady state limit ($t>>1/\gamma$) we are interested
in the particular solution of the set of equations of motion 
of $\xi$. We use the Fourier transform of
\begin{equation}
\xi_j(t)=\int_{-\infty}^\infty\frac{d\omega}{2\pi}\xi_j(\omega)
{\rm e}^{i\omega t}
~{\rm and}~
f_j(t)=\int_{-\infty}^\infty\frac{d\omega}{2\pi}f_j(\omega)
{\rm e}^{i\omega t}
\label{ftxif}
\end{equation}
in Eqn.(\ref{eomxi}) and obtain
\begin{equation}
(-\omega^2\delta_{jk}+\alpha_j^2\delta_{jk}+i\gamma\omega Z_{jk})
\xi_k(\omega)=\tilde{f}_j(\omega).
\label{eomxiomega}
\end{equation}
Since the dynamics of the system in the steady state is governed
by the noises, we decompose $\xi_j(\omega)$ as
\begin{equation}
\xi_j(\omega)=b(\omega)\tilde{f}_j(\omega)
\label{xiomega}
\end{equation} 
and then using this decomposition into Eqn.(\ref{eomxiomega}) we 
obtain
\begin{equation}
b(\omega) =
-\frac{1}{\omega^2-\alpha_j^2-i\gamma\omega}.
\label{b}
\end{equation}
Now upon substitution of Eqn.(\ref{xiomega}) into (\ref{ftxif}) 
along with the use of Eqn.(\ref{fj}), (\ref{b})  
we obtain
\begin{eqnarray}
\xi_j(p_2,p_3,t) &=& -\int_{-\infty}^\infty\frac{d\omega}{2\pi}
\frac{{\rm e}^{i\omega t}}{\omega^2-\alpha_j^2-i\gamma\omega}\nonumber\\
& &\times[a^{(j)}_1\,f_L(p_2,p_3,\omega)
+a^{(j)}_N\,f_R(p_2,p_3,\omega)].\nonumber\\
& &
\label{xiss}
\end{eqnarray}
Now the use of Eqn.(\ref{ftetan}), (\ref{fj}) and (\ref{ftxif}) 
into (\ref{noise1}) gives
\begin{eqnarray}
& &\langle f_{L,R}(p_2,p_3,\omega)f_{L,R}(p_2^\prime,p_3^\prime,\omega^\prime)
\rangle\nonumber\\
&=&4\pi\gamma T_{L,R}\,\delta(\omega+\omega^\prime)\,\delta_{p_2+p_2^\prime,0}
\,\delta_{p_3+p_3^\prime,0}.
\label{noise2}
\end{eqnarray}
To compute the correlation between position and velocity
we use Eqn.(\ref{xiss}) and (\ref{noise2}) and after performing 
a frequency integration using delta function obtain
\begin{eqnarray}
& &\langle \xi_{k_1}(p_2,p_3,t)\dot\xi_{k_2}(p_2^{\prime},
p_3^{\prime},t^\prime)\rangle \nonumber\\
&=&2\gamma
\,(a_1^{(k_1)}a_1^{(k_2)}T_L+a_N^{(k_1)}a_N^{(k_2)}T_R)
\,I_c(t-t^\prime)\nonumber\\
& &\times\delta_{p_2+p_2^{\prime},0}
\,\delta_{p_3+p_3^{\prime},0},
\label{correlation}
\end{eqnarray}
where
\begin{eqnarray}
& &I_c(t-t^\prime)\nonumber\\
&=& -i\int_{-\infty}^\infty\frac{d\omega}{2\pi}
\frac{\omega {\rm e}^{i\omega(t-t^\prime)}}
{(\omega^2-\alpha_{k_1}^2-i\gamma\omega)
(\omega^2-\alpha_{k_2}^2+i\gamma\omega)}.
\end{eqnarray}  
Performing the integration over $\omega$ we obtain
\begin{eqnarray}
I_c(t-t^\prime) &=& \frac{{\rm e}^{-\gamma|t-t^\prime|/2}}
{4\,\Delta_d(\beta_1,\beta_2)}[I_c^>(t-t^\prime)\theta(t-t^\prime)\nonumber\\
& &+\,I_c^<(t-t^\prime)\theta(t^\prime-t)],
\label{ic}
\end{eqnarray}
where
\begin{eqnarray}
\Delta_d(\beta_1,\beta_2) &=& (\cos\beta_1-\cos\beta_2)^2\nonumber\\
& &+\,\gamma^2(2\omega_0^2
+\cos\beta_1+\cos\beta_2),\label{deld}\\
I_c^>(t-t^\prime) &=& 2(\cos\beta_1-\cos\beta_2)
\cos(\omega_{k_1}|t-t^\prime|)\nonumber\\
& &+\,\frac{\gamma}{\omega_{k_1}}\{(4\omega_0^2+3\cos\beta_1
+\cos\beta_2)\nonumber\\
& &\times\sin(\omega_{k_1}|t-t^\prime|)\},\label{icg}\\
I_c^<(t-t^\prime) &=& 2(\cos\beta_1-\cos\beta_2)
\cos(\omega_{k_2}|t-t^\prime|)\nonumber\\
& &-\,\frac{\gamma}{\omega_{k_2}}\{(4\omega_0^2+\cos\beta_1
+3\cos\beta_2)\nonumber\\
& &\times\sin(\omega_{k_2}|t-t^\prime|)\},\label{icl}\\
\beta_{1,2} &=& \pi k_{1,2}/(N+1),\label{beta12}\\
\omega_{k_{1,2}} &=& \sqrt{\alpha^2_{k_{1,2}}-\gamma^2/4}.
\label{omegak12}
\end{eqnarray}
It is clear that $I_c(t-t^\prime)\rightarrow 0$, when
$|t-t^\prime|\rightarrow\infty$ and when $t=t^\prime$
\begin{equation}
I_c(0) = \frac{\cos\beta_1-\cos\beta_2}{2\,\Delta_d(\beta_1,\beta_2)}.
\label{ic0}
\end{equation}
For $1\le|k_1-k_2|\le N-1$, $I_c(0)$ remains finite when $N$ tends 
to infinity. According to Eqn.(\ref{evec}) the factor  
appeared in Eqn.(\ref{correlation})
$(a_1^{(k_1)}a_1^{(k_2)}T_L+a_N^{(k_1)}a_N^{(k_2)}T_R)
=2(T_L+(-1)^{k_1+k_2}T_R)\sin\beta_1\sin\beta_2/(N+1)$. It
implies that even for zero momentum modes ($p_{2,3}=0$),
which appear owing to the periodic boudary conditions imposed
on the displacement field in $n_2$ and $n_3$ directions,
the equal time correlation in Eqn.(\ref{correlation})
goes as $N^{-\alpha}$ ($1\le\alpha\le 3$) when $N\rightarrow\infty$.
The fall of this correlation as a negative power of $N$ 
in the thermodynamic limit indicates that the ballastic transport 
remains absent from the conduction process of heat\cite{Grassberger2}.

\begin{figure}[!th]
\begin{center}
\includegraphics*[width=0.48\textwidth,clip=true]{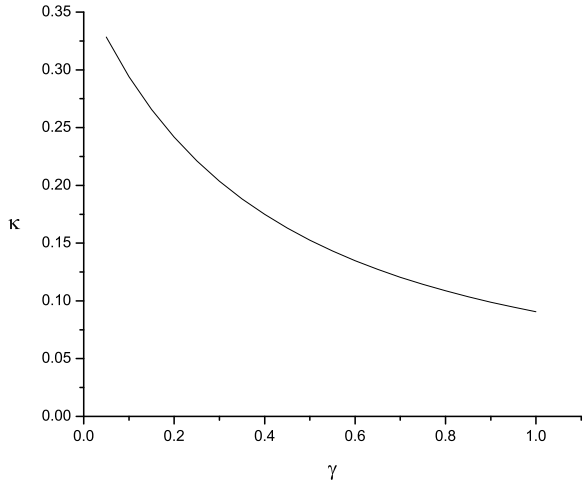}
\end{center}
\vskip -0.45in
\caption[]{(Color online) Plot of $\kappa$ as a function of $\gamma$} 
\label{Figkg}
\end{figure}

Heat current density $j_{\bf n}$ from the 
lattice site ${\bf n}$ to
${\bf n+\hat{e}_1}$, where ${\bf \hat{e}_1}=(1,0,0)$, 
is given by\cite{Lepri1}
\begin{eqnarray}
j_{\bf n}=\frac{1}{2}\langle (x_{\bf n + \hat{e}_1}-x_{\bf n})
(\dot{x}_{\bf n + \hat{e}_1}+\dot{x}_{\bf n})\rangle
\end{eqnarray}
The average heat current density per bond\cite{Dhar4} 
\begin{equation}
J = \frac{1}{2W_2W_3 (N-1)}
\sum_{n_1=1}^{N-1}\sum_{n_2=1}^{W_2}\sum_{n_3=1}^{W_3} j_{\bf n}.
\end{equation}
We substitute Eqn.(\ref{ftxn}) and (\ref{yxi}) in $J$ and after 
performing the summations over $n_2$ and $n_3$ obtain the 
average heat current density per bond in the steady state limit as
\begin{eqnarray}
J&=&\frac{1}{2W_2W_3 (N-1)}
\,\sum_{p_2,p_3}\, \sum_{k_1,k_2=1}^N
\,\sum_{n_1=1}^{N-1}(a_{n_1+1}^{(k_1)} - a_{n_1}^{(k_1)})\nonumber\\
& &\times(a_{n_1+1}^{(k_2)} + a_{n_1}^{(k_2)})
\langle \xi_{k_1}(p_2,p_3,t)\dot{\xi}_{k_2}(-p_2,-p_3,t)\rangle.\nonumber\\
& &
\end{eqnarray}
We now use Eqn.(\ref{evec}) to evaluate the sum
\begin{eqnarray}
& &\sum_{n_1=1}^{N-1}(a_{n_1+1}^{(k_1)} - a_{n_1}^{(k_1)})
(a_{n_1+1}^{(k_2)} + a_{n_1}^{(k_2)})\nonumber\\
&=&2(1-(-1)^{k_1+k_2})\sin\beta_1\sin\beta_2\nonumber\\
& &\times\Big[\frac{1}{\cos\beta_2-\cos\beta_1}-1\Big]
\label{n1sum}
\end{eqnarray}
and then using (\ref{correlation}) and (\ref{ic0}) obtain
\begin{eqnarray}
J&=&-\frac{2\gamma\,(T_L-T_R)}
{(N+1)^2(N-1)W_2W_3}
\,\sum_{p_2,p_3}\, \sum_{k_1,k_2=1}^N\nonumber\\ 
& &\times(1-(-1)^{k_1+k_2})
\frac{\sin^2\beta_1 \sin^2\beta_2}
{\Delta_d(\beta_1,\beta_2)}. 
\label{hc1}
\end{eqnarray}\\
The factor $(1-(-1)^{k_1+k_2})$ ensures that the summation 
over $k_1$ and  $k_2$ will be non zero only when $k_1+k_2$ 
is an odd number and hence we take the   
factor $(T_L+(-1)^{k_1+k_2}T_R)$ 
out of the summation as $(T_L-T_R)$.
In the continuum limit, when $a\rightarrow 0$ and 
$W_{2,3}\rightarrow \infty$ keeping $a\,W_{2,3}$ at fixed
values, we convert the discrete sums over $p_2$ and $p_3$
into integrals:
\begin{equation}
\sum_{p_{2,3}} \rightarrow \frac{a\,W_{2,3}}{2 \pi}
\int_{-\frac{\pi}{a}}^{-\frac{\pi}{a}}\,dp_{2,3}.
\end{equation}
Evaluation of the integrals\cite{Gradshteyn} over $p_2$ and $p_3$ 
gives
\begin{equation}
J=-\frac{2\gamma\,(T_L-T_R)}{N-1} I(N,\gamma),
\end{equation}
where
\begin{eqnarray}
& &I(N,\gamma) = \frac{1}{(N+1)^2}\sum_{k_1,k_2=1}^N
(1-(-1)^{k_1+k_2})\nonumber\\
& &\times\frac{\sin^2\beta_1\sin^2\beta_2}{\Delta(\beta_1,\beta_2)} 
F\left(\frac{1}{2},\frac{1}{2},1;
\,(4\gamma^2/\Delta(\beta_1,\beta_2))^2\right).
\label{ing}
\end{eqnarray}
Here the function
\begin{eqnarray}
\Delta(\beta_1,\beta_2) &=& (\cos\beta_1-\cos\beta_2)^2\nonumber\\
& &\,+\gamma^2\,(6+\cos\beta_1+\cos\beta_2).
\label{delta}
\end{eqnarray}
$I(N,\gamma)$ is zero if $k_1$ and $k_2$ simultaneously take even
integer values or odd integer values. Assuming that $N$ be an even
number and using the fact that the summand of Eqn.(\ref{ing}) is
symmetric in respect of the interchange of $\beta_1$ and $\beta_2$, 
we rewrite the double sum of
\begin{eqnarray}
I(N,\gamma) &=& \frac{4}{(N+1)^2}\sum_{j_1,j_2=1}^{N/2}
\frac{\sin^2\tilde{\beta}_1\sin^2\tilde{\beta}_2}
{\Delta(\tilde{\beta}_1,\tilde{\beta}_2)}\nonumber\\ 
& &\times F\left(\frac{1}{2},\frac{1}{2},1;
\,(4\gamma^2/\Delta(\tilde{\beta}_1,\tilde{\beta}_2))^2\right),
\label{ing1}
\end{eqnarray}
where $\tilde\beta_1=2\pi j_1/(N+1)$ 
and $\tilde\beta_2=\pi(2j_2-1)/(N+1)$. 
Again in the continuum limit we convert this double sum into integrals.  
In this limit $a\rightarrow 0$ and $N\rightarrow\infty$
keeping $Na$ at a fixed value. Defining the integration variables
in this limit as $\theta_{1,2}=2\pi j_{1,2}/(N+1)$, we convert
the discrete sums into integrals:
\begin{equation}
\frac{2}{N+1}\sum_{j_{1,2}=1}^{N/2}\rightarrow
\frac{1}{\pi}\int_0^\pi d\theta_{1,2}.
\end{equation} 
$I(N,\gamma)$ thus takes the form
\begin{eqnarray}
g(\gamma) &=& \lim_{N\rightarrow\infty}I(N,\gamma)\nonumber\\
&=&\frac{1}{\pi^2}\int_0^\pi d\theta_1\int_0^\pi d\theta_2
\frac{\sin^2\theta_1\sin^2\theta_2}
{\Delta(\theta_1,\theta_2)}\nonumber\\ 
& &\times F\left(\frac{1}{2},\frac{1}{2},1;
\,(4\gamma^2/\Delta(\theta_1,\theta_2))^2\right).
\label{gg}
\end{eqnarray}
Hence we obtain the steady state current density per bond in the
continuum limit 
\begin{equation}
J = -\kappa\frac{(T_L-T_R)}{N-1},
\end{equation}
where the conductivity 
\begin{equation}
\kappa=2\gamma\,g(\gamma).
\label{kpgm}
\end{equation}
Here $\kappa$ is found to be independent of the size of the system. 
The variation of the thermal conductivity $\kappa$
as a function of $\gamma$, as given by Eqn.(\ref{kpgm}),
is plotted in Fig.\ref{Figkg}.
Here $\gamma$ appears as a constant in the dissipative force 
term of the Langevin's equation. 
Physically this force term denotes a viscous force experienced by the
particles of Brownian like at the boundary surfaces of the
crystal owing to collisions with the particles of fluid 
which seems to constitute the heat baths\cite{Chaikin}. 
The increase of $\gamma$, reduces the mobilities of the
Brownian particles and thereby reducing their 
velocities\cite{Chaikin,Reif}. Consequently, the
velocities of the particles at the surfaces next to the
boundaries will also fall because those are connected by
springs with the particles at the boundaries. This fall
of velocities of the particles at the neighbouring surfaces
of the boundaries will reduce the rate of flow of heat from
the boundaries to the crystal itself and thereby reducing 
the thermal conductivity of the system. Hence, it justifies 
reasonably the nature of variation of $\kappa$ with $\gamma$ 
as shown in Fig.\ref{Figkg}.

The average of the square of velocity
of a layer at $n_1$ reads 
\begin{eqnarray}
v_{avg}^2(n_1) &=& \frac{1}{W_2W_3}
\sum_{n_2=1}^{W_2}\sum_{n_3=1}^{W_3}
\langle{\dot{x}}^2_{\bf{n}}\rangle\nonumber\\
&=&\frac{1}{W_2W_3}\sum_{p_2,p_3}\sum_{k_1,k_2=1}^N
a_{n_1}^{(k_1)}a_{n_1}^{(k_2)}\nonumber\\
& &\times\langle\dot{\xi}_{k_1}(p_2,p_3,t)
\dot{\xi}_{k_2}(-p_2,-p_3,t)\rangle.
\label{vav1}
\end{eqnarray}
We use Eqn.(\ref{xiss}) to compute the velocity-velocity correlation as
\begin{eqnarray}
& &\langle\dot{\xi}_{k_1}(p_2,p_3,t)
\dot{\xi}_{k_2}(-p_2,-p_3,t)\rangle \nonumber\\
&=& \frac{2\gamma^2}{N+1}(T_L+(-1)^{k_1+k_2})
\sin\beta_1\sin\beta_2\nonumber\\
& &\times\frac{2\omega_0^2+\cos\beta_1+\cos\beta_2}
{\Delta_d(\beta_1,\beta_2)}
\label{vvcorrelation}
\end{eqnarray}
Upon substitution of Eqn.(\ref{vvcorrelation}) into 
Eqn.(\ref{vav1}) and evaluation of $p_2$ and $p_3$ sum
in the continuum limit along $n_2$ and $n_3$ directions, give
\begin{equation}
v_{avg}^2(n_1)=h_L(n_1,N)T_L+h_R(n_1,N)T_R
\end{equation}
where
\begin{eqnarray}
h_L(n_1,N) &=& \frac{4}{(N+1)^2}\sum_{k_1,k_2=1}^N
\frac{\Lambda(\beta_1,\beta_2)}{\Delta(\beta_1,\beta_2)}\nonumber\\
& &\times\sin(n_1\beta_1)\sin(n_1\beta_2)\sin\beta_1\sin\beta_2,
\label{hl}\\
h_R(n_1,N) &=& \frac{4}{(N+1)^2}\sum_{k_1,k_2=1}^N
(-1)^{k_1+k_2}\frac{\Lambda(\beta_1,\beta_2)}
{\Delta(\beta_1,\beta_2)}\nonumber\\
& &\times\sin(n_1\beta_1)\sin(n_1\beta_2)\sin\beta_1\sin\beta_2
\label{hr},\\
\Lambda(\beta_1,\beta_2) &=& \{(\cos\beta_1-\cos\beta_2)^2\nonumber\\
& &\times[1- F(1/2, 1/2, 1; (4\gamma^2/\Delta(\beta_1,\beta_2))^2)]\}
\nonumber\\
& &+\gamma^2(6+\cos\beta_1+\cos\beta_2).
\label{bigl}
\end{eqnarray}
\begin{figure}[!th]
\begin{center}
\includegraphics*[width=0.48\textwidth,clip=true]{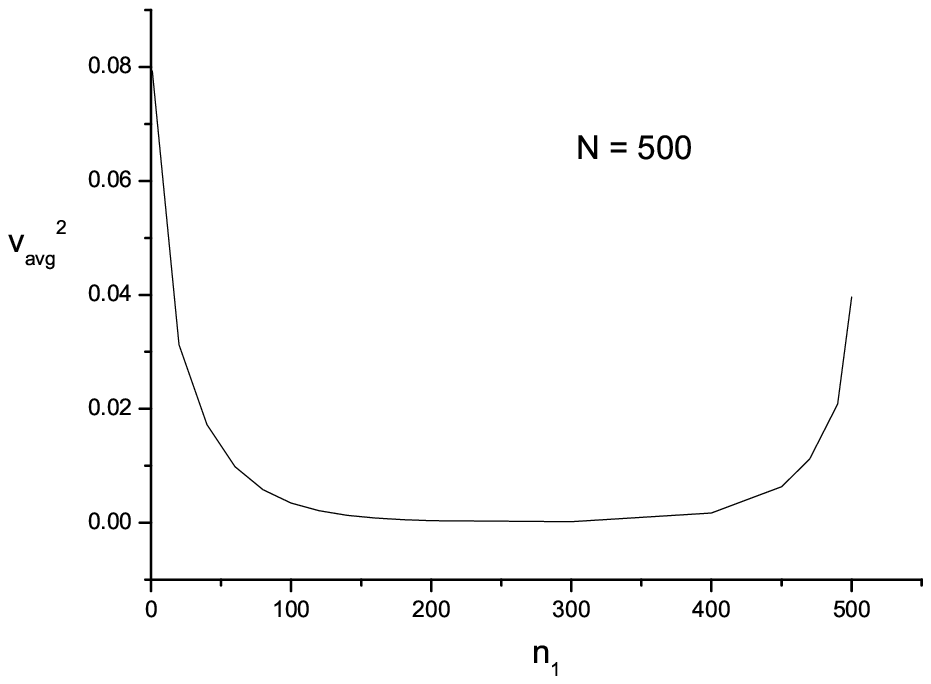}
\end{center}
\vskip -0.45in
\caption[]{(Color online) Plot of $v_{avg}^2$ as a function of $n_1$}
\label{Figavg1}
\end{figure}
\begin{figure}[!th]
\begin{center}
\includegraphics*[width=0.48\textwidth,clip=true]{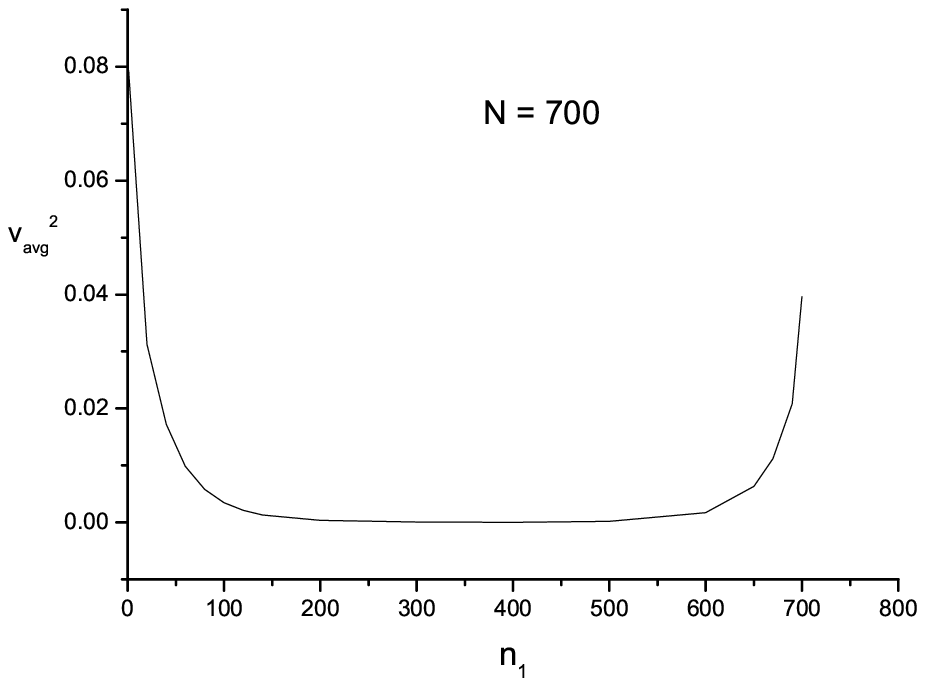}
\end{center}
\vskip -0.45in
\caption[]{(Color online) Plot of $v_{avg}^2$ as a function of $n_1$}
\label{Figavg2}
\end{figure}
Our evaluation suggests that for $\gamma=0.01$, $h_L$ tends 
to $0.0396$ and $0$ and $h_R$
tends to $0$ and $0.0396$ at $n_1=1$ and $n_1=N$ respectively
when $N\rightarrow\infty$. It indicates that as
$h_L$ and $h_R$ are monotonically
decreasing and increasing functions of $n_1$ respectively, 
$v_{avg}^2$ attains a minimum at any layer in the region between 
$n_1=1$ and $n_1=N$ and it is also evident from our plots
given in Fig.\ref{Figavg1} and \ref{Figavg2}. 
Since, $v_{avg}^2(n_1)$ is 
proportional to $T(n_1)$, the temperature of the layer at $n_1$, 
$T(n_1)$ also  exhibits a minimum in the region
$1<n_1<N$. This concave upward nature of $T(n_1)$  
has also been predicted in Ref.\cite{Dhar4}
 
In summary, we have given an exact analytical derivation of
Fourier's law of heat conduction in a three dimensional
harmonic crystal. It shows that in three dimensions
without introducing any pinning or disorder, harmonicity
alone can give rise to a normal transport of heat in the 
crystal in the continuum limit.

\end{document}